\documentclass[letterpaper, 10 pt, conference]{ieeeconf}  

\IEEEoverridecommandlockouts                              

\usepackage{graphicx} 
\usepackage{xcolor}
\usepackage{amsmath}
\usepackage{amssymb}
\usepackage[ruled,vlined]{algorithm2e}
\usepackage{multirow}
\usepackage{multicol}
\usepackage{caption}
\usepackage[space, nocompress]{cite}
 \usepackage{tabularx}
 \usepackage{tikz}
\usepackage{textcomp}
\usepackage{hyperref}
\usepackage{lipsum}
\usepackage{fancyhdr}

\title{\LARGE \bf
Multimodal Fusion Using Deep Learning Applied to Driver's Referencing of Outside-Vehicle Objects}

\author{Abdul Rafey Aftab$^{1,2}$, Michael von der Beeck$^1$, Steven Rohrhirsch$^1$, Benoit Diotte$^1$, Michael Feld$^2$
\thanks{$^{1}$Abdul Rafey Aftab, Michael von der Beeck, Steven Rohrhirsch and Benoit Diotte are with BMW Group, Munich, Germany 
{\tt\scriptsize \{abdul-rafey.aftab, michael.beeck, steven.rohrhirsch, benoit.diotte\}@bmw.de}}%
\thanks{$^{2}$Michael Feld is with German Research Center for Artificial Intelligence (DFKI), Saarbr\"ucken, Germany
{\tt\scriptsize michael.feld@dfki.de}}%
}

\newcommand\copyrighttext{%
  \footnotesize \textcopyright 2021 IEEE. Personal use of this material is permitted. Permission from IEEE must be obtained for all other uses, in any current or future media, including reprinting/republishing this material for advertising or promotional purposes, creating new collective works, for resale or redistribution to servers or 
  lists, or reuse of any copyrighted component of this work in other works. 
  }
\newcommand\copyrightnotice{%
\begin{tikzpicture}[remember picture,overlay]
\node[anchor=south,yshift=10pt] at (current page.south) {\fbox{\parbox{\dimexpr\textwidth-\fboxsep-\fboxrule\relax}{\copyrighttext}}};
\end{tikzpicture}%
}

\begin{document}

\maketitle
\copyrightnotice

\thispagestyle{fancy}
\pagestyle{plain}

\begin{abstract}

There is a growing interest in more intelligent natural user interaction with the car. Hand gestures and speech are already being applied for driver-car interaction. Moreover, multimodal approaches are also showing promise in the automotive industry. In this paper, we utilize deep learning for a multimodal fusion network for referencing objects outside the vehicle. We use features from gaze, head pose and finger pointing simultaneously to precisely predict the referenced objects in different car poses. We demonstrate the practical limitations of each modality when used for a natural form of referencing, specifically inside the car. As evident from our results, we overcome the modality specific limitations, to a large extent, by the addition of other modalities. This work highlights the importance of multimodal sensing, especially when moving towards natural user interaction.  Furthermore, our user based analysis shows noteworthy differences in recognition of user behavior depending upon the vehicle pose. 

\end{abstract}


\section{INTRODUCTION}

In the automotive industry, user-centered natural interaction inside the car is gaining prominence. The in-vehicle functions, mainly used for secondary tasks, affect the driver's cognitive attentiveness as they require off-road eye glances or touch presses \cite{klauer2010analysis}. Thus, there is a need for natural forms of interaction with the vehicle to decrease distractions. Speech based voice assistants overcome the off-road glances, but can be mentally tedious when describing deictic references. By using hand gestures and eye movements with speech, in addition to reduced visual demand \cite{riener2012gestural, manawadu2017multimodal, rumelin2013free}, there is the added advantage of having naturalness and increased simplicity, as compared to touch based interaction \cite{tscharn2017stop}. These multimodal interactions, starting from Bolt's seminal work \cite{bolt1980put}, have been studied and incorporated inside the car by previous researchers \cite{ohn2014head, nesselrath2016combining, gomaa2020studying}.

While the integration of multiple modalities has the potential to outperform mono-modal systems \cite{turk2014multimodal, esteban2005review}, many current research methods rely on using single input modalities or using other modalities in a passive manner, such as a trigger \cite{schweigert2019eyepointing, chatterjee2015gaze+, vora2017generalizing, gomaa2020studying}. On the one hand, eye gaze imposes limitations with the always-on characteristic and is very volatile. On the other hand, mid-air gestures for referencing can be subjective and result in discrepancies among users. It is important to mention here that, within the context of this paper, the type of gesture we deal with is a deictic gesture by pointing with a finger, which the users do not need to learn. It has been observed that drivers are relatively imprecise in pointing \cite{brand2016pointing, roider2018implementation}, and the integration of gaze improves the accuracy of pointing while driving \cite{roider2018see}.

\begin{figure}[t]
    \centering
    \includegraphics[trim=0 0 0 0, clip, width=.485\textwidth]{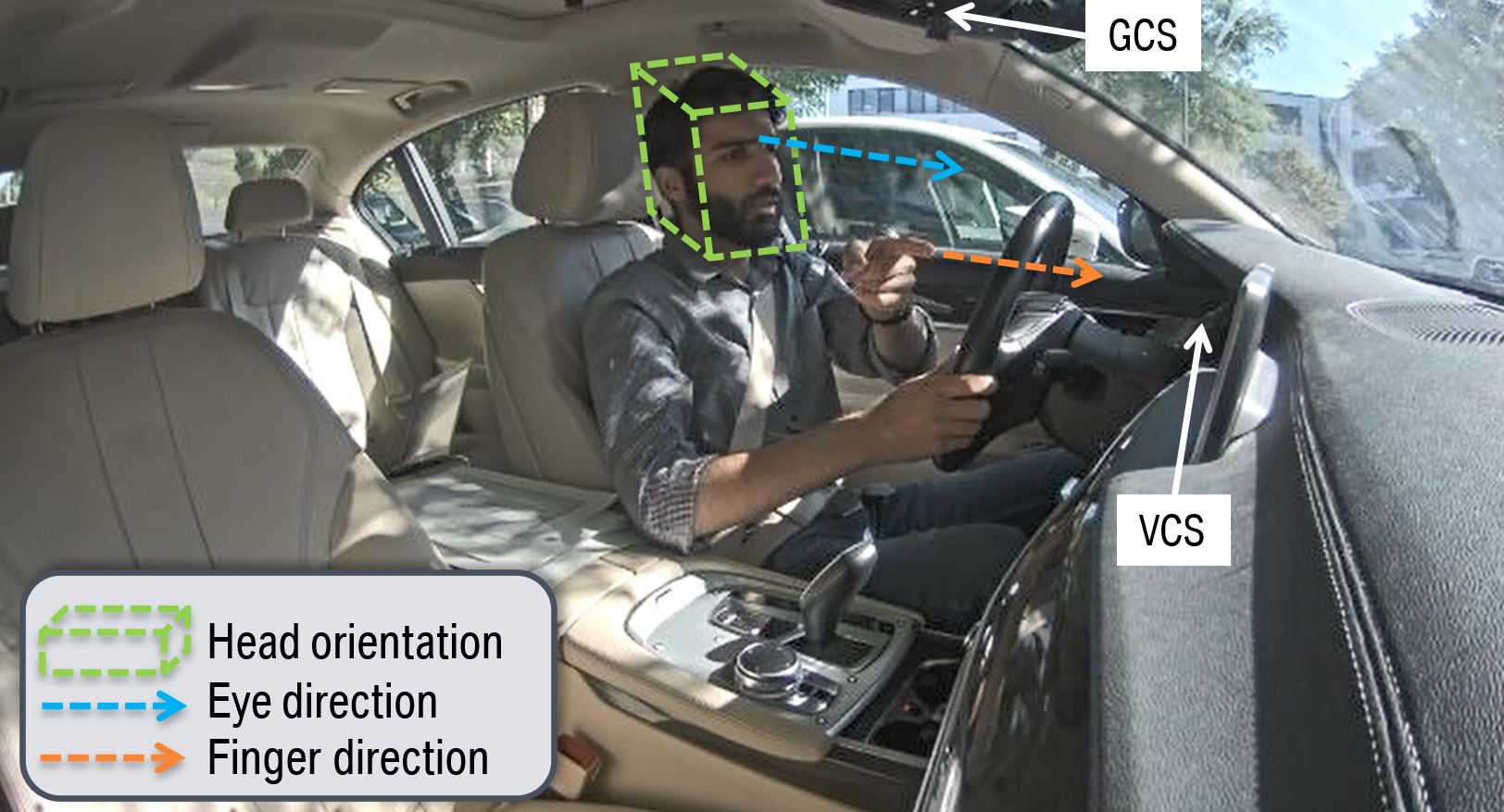} \\
    \captionof{figure}{Driver points to the right side with the left hand, occluding his face from gaze and head tracker.}
    \label{fig:teaser}
\end{figure}

In this work, we integrate the deictic information from gaze, head pose and finger pointing gesture in order to reference outside vehicle objects. This multimodal integration combines the efficiency and naturalness of these modes of interaction by overcoming the drawbacks of a single modality using others and enhances recognition performance by taking advantage of the temporal relations between modalities. To this end, we implement a deep model-level fusion neural network by concatenating high-level feature representations of modalities \cite{chen2016multi, wu2014survey}. This research extends our previous work \cite{aftab2020you}, where we use the three modalities for selection of control modules inside the vehicle, to a new application. In this sequel, the driver can freely refer to landmarks and, with the help of advanced sensing and machine learning, the car  should be able to understand the driver's referenced object. Recognition of the action to perform is a different topic which we do not explore. Examples of such natural interaction are inquiries about a certain building, such as "what kind of restaurant is \textit{that}?", or commands like "stop over \textit{there}" \cite{tscharn2017stop}. An application of such a multimodal interaction was presented at the Mobile World Congress 2019, namely the BMW Natural Interaction \cite{bmwnatint}. 

\section{Related Work}
Ample research has been conducted on the use of eye gaze, head pose or gestures for user interaction. Eye tracking is often applied to control a cursor on screen, such as in MAGIC pointing \cite{zhai1999manual} and its extension EyePointing \cite{schweigert2019eyepointing},  which use a manual trigger and a finger pointing gesture for selection, respectively. Kang \textit{et. al.} designed a gazed based driver query prototype using where the car is able to predict the queried building with an accuracy of 65\% \cite{kang2015you}. Nevertheless, selection exclusively by gaze  input is cumbersome, especially when objects are placed close to each other \cite{hild2019suggesting}. The use of free hand pointing provides naturalness as well as reduced cognitive demand while driving \cite{rumelin2013free, tscharn2017stop}. Gomaa \textit{et. al.} study user behavior with gaze and finger pointing in a driving simulator and show gaze accuracy to be better than pointing accuracy \cite{gomaa2020studying}. 

A multimodal approach using gaze, speech and gestures to select objects is presented by Nesselrath \textit{et al.} \cite{nesselrath2016combining}. The integration of gestures and gaze as inputs has been demonstrated to have better outcomes as compared to systems with either gesture or gaze  \cite{chatterjee2015gaze+}. These approaches use gaze tracking as a primary source of input and enhance the naturalness with a secondary input such as gestures. However, they lack the enhancement in the preciseness of gaze tracking. Unlike these, we use multimodal fusion for better precision of the driver's referenced direction as head, gaze and finger pointing all provide relevant deictic information \cite{akkil2016accuracy}.

Using finger pointing to recognize objects that do not lie straight ahead is a challenging task \cite{akkil2016accuracy}. To improve the pointing gesture accuracy for object selection inside the car, Roider {et. al.} \cite{roider2018see} use a simple rule based approach involving gaze tracking. In our prior work, we tackled the problem of accurately selecting control modules in the cockpit of the vehicle using a CNN based fusion model \cite{aftab2020you}. We used head pose simultaneously with gaze and finger pointing, because the head pose and gaze direction are directly related and usually considered  together for recognizing the visual behavior \cite{ji2002real, ghosh2020speak2label}. Head pose typically provides better availability, whereas gaze reveals better preciseness when the driver focuses on an object. 

Motivated by applications of deep neural networks for multimodal fusion in the past \cite{ngiam2011multimodal, meng2020survey, chen2016multi}, we apply deep Convolutional Neural Networks (CNN) on the preprocessed sequential (temporal) input data. Each modality feature is treated equally at the input of the  fusion network, and the network learns the weight from the training data. Furthermore, with the model-level fusion, the temporal relations between modalities can be exploited \cite{wu2014survey}.

To summarize, majority of the studies are performed in a simulated environment with added limitations, such as pointing only to a screen or using only right hand. We perform our experiments in a real car with an authentic environment to have more realistic results than a simulation. Furthermore, we collected our data from different angles to add a larger variation in the pointing direction. Even though driver's gaze to query objects outside the vehicle has been studied extensively, simultaneous use of finger pointing with gaze has not. One other aspect which the state-of-the-art lacks is the viability of the non-contact sensors. Non-contact sensors for eye and gesture tracking impose certain constraints such as occlusion or out of field-of-view. In our study, we impose minimal limitations to allow users to behave naturally, and we discuss the problems that arise with a practical experimental setup in real-world situations. 

\begin{figure*}[t]
    \centering
    \includegraphics[trim=0 0 0 0,clip, width=\linewidth]{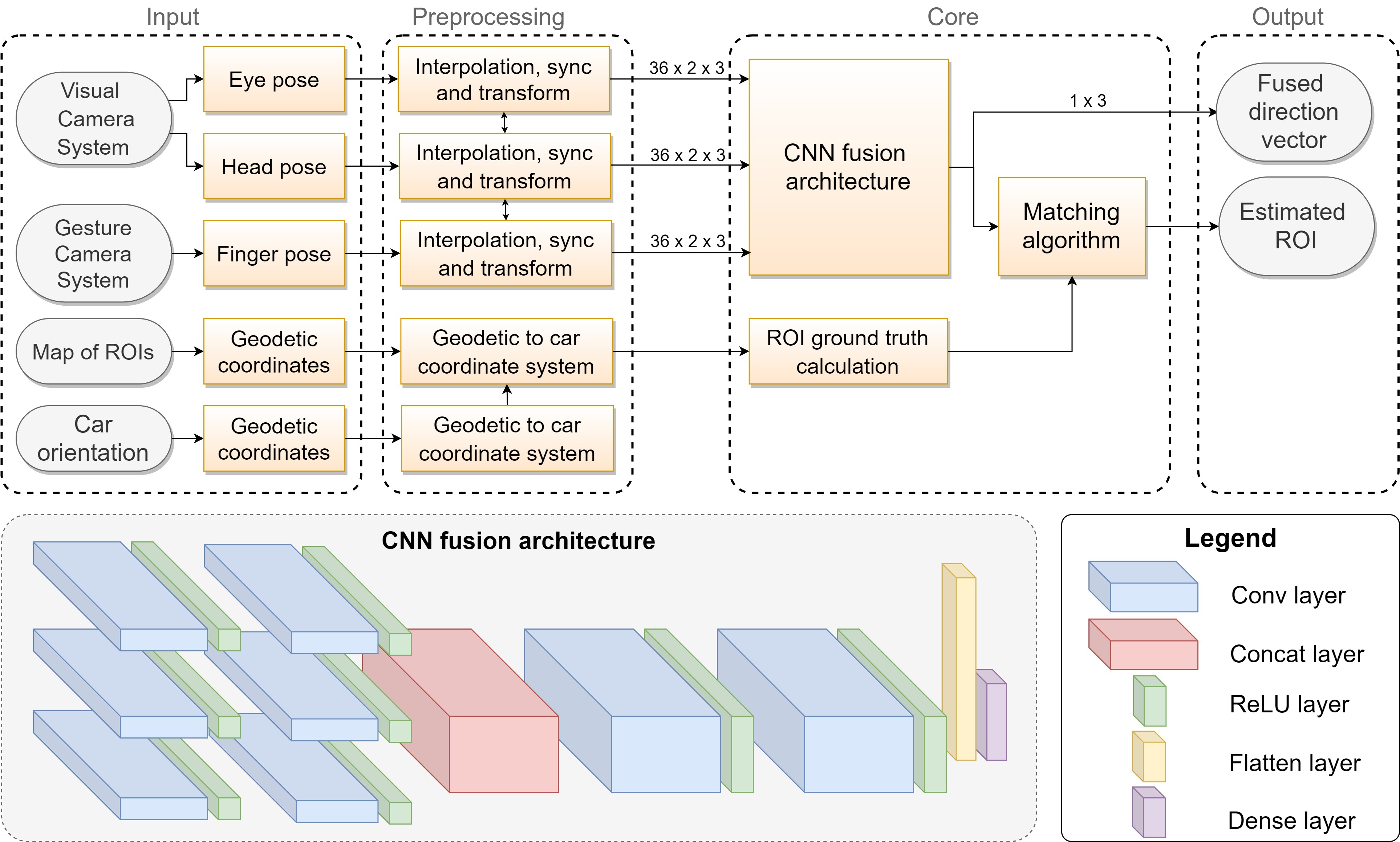}
    \caption{A multimodal model-level fusion architecture using a deep CNN.}
    \label{fig:methodology}
\end{figure*}

\section{Experimental Setup}
 For data collections, a functional vehicle was used in different locations to have a variety in pointing directions. However, all data was collected in a stationary car to mimic the self-driving cars of the future. This made collecting ground truth data relatively easier as compared to a car being driven. While driving, the primary focus of the driver is on the road \cite{rumelin2013free}. In contrast, free-hand pointing from a stationary car would primarily be focused on the pointing task, which allows drivers to be relatively more composed. To not interfere with the driver's comfort, non-contact sensors were utilized in a non-intrusive way.

\subsection{Apparatus}
The apparatus used is similar to the one in \cite{aftab2020you}. Two types of non-commercial prototype cameras were used to capture the driver's gestural and visual features, referred to as the Gesture Camera System (GCS) and the Visual Camera System (VCS) respectively. Additionally, four other cameras were installed in the car to record the events inside the car. 

\subsubsection{GCS} 
The gesture camera was mounted at the car ceiling adjacent to the rear-view mirror (see Figure \ref{fig:teaser}). It captured the hand and finger movements in the 3D space using a Time-of-Flight camera with a wide angle Field-of-View (FoV). The gesture camera system detects a finger pointing gesture and calculates the vector from the tip of the finger to the base of the finger. The 3D coordinates of the fingertip are used as the finger position.

\subsubsection{VCS}
The visual camera system, installed behind the steering wheel, evaluates the images of the driver for the head and eye poses (where pose constitutes both the position and direction). The algorithm integrated into the camera system calculates the head orientation as three Euler angles (roll, pitch and yaw), and the 3D coordinates of the estimated centre position of the head. In addition to this, it provides eye position as the 3D coordinates of the centre point between the two eyes, and the 3D vector coordinates of the eye direction merged together from both the left and the right eye (shown in Figure \ref{fig:teaser}).

\subsubsection{Speech}
Along with the pointing gesture, a specific speech command was used as a trigger which we implemented with the Wizard-of-Oz (WoZ) method. To record the timestamp of the pointing gesture, a secondary person (acting as the wizard) pressed a button manually when the primary user (i.e. driver) pointed towards an object and said, "what is \textit{that}?", such that the timestamp is noted at the instant when "\textit{that}" is said. However, it may involve human error. 

\subsection{Feature Extraction and Preprocessing} \label{sec:feats}

The out-of-box gestural and visual features from both the camera systems included:
\begin{itemize}
    \item Finger pose: the position and (normalized) direction vector  in the 3D vector space of the finger.
    \item Eye pose: the position and (normalized) direction vector in the 3D vector space of the eye gaze.
    \item Head pose: the 3D position of the centre of the head and the Euler angles (in radians) of the head orientation.
\end{itemize}

In some of the referencing events, due to occlusion, there were frames with missing features. These mainly occurred when the arm came in front of face (as in Figure \ref{fig:teaser}) or when the head is turned to the far sides so that eyes can no longer be tracked. In some cases, the participants extended their arms beyond the field-of-view of the gesture camera, especially when pointing with the left hand to the left side. To fill the gaps in the missing frames, we used linear interpolation from the two nearest neighbouring frames. 

\begin{figure}[t]
    \centering
    \includegraphics[trim=0 0 0 0,clip, width=\linewidth]{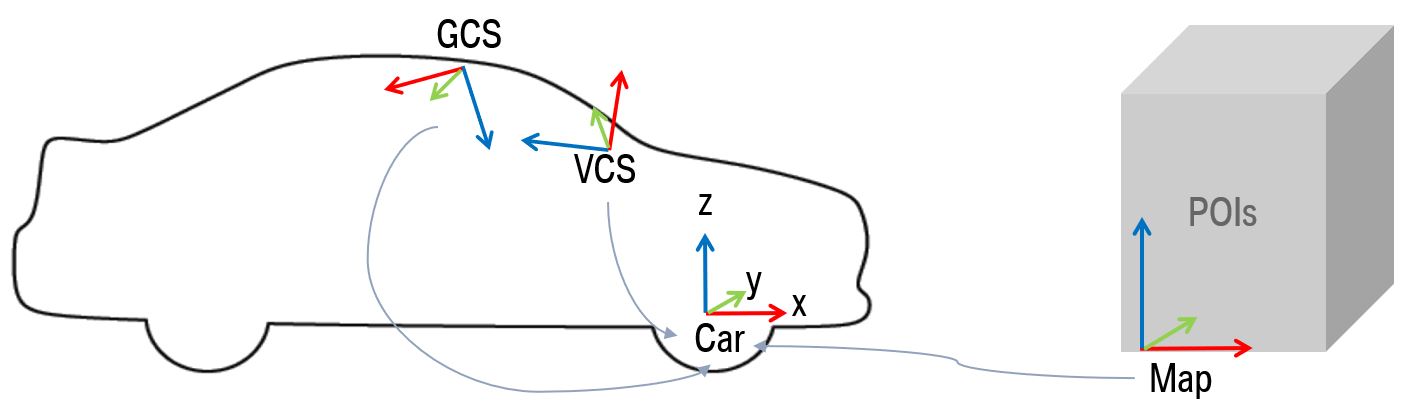}
    \caption{Camera systems and map with their coordinate system are transformed to the car coordinate system.}
    \label{fig:sensors}
\end{figure}

The out-of-the-box features were in their respective camera coordinate system. Rotation and translation matrices for both cameras were calculated using markers on the windshield, and the features were transformed to the car coordinate system following the ISO 8855 \cite{iso2011road} standard with the origin at the centre of the front car axle (see Figure \ref{fig:sensors}). 

Finally, we selected a time interval of 0.8 seconds as input, which is observed to be a comfortable pointing time \cite{rumelin2013free}. The time interval of 0.8 seconds was chosen such that it would have 0.4 seconds before the noted timestamp from the WoZ button press and 0.4 seconds after it.

\subsection{Participants}
The experiment involved 28 participants out of which 3 were female, aged between 23 and 57 years old (mean 32 years old). 7 of the participants wore glasses, 2 wore contact lenses, and the remaining did not wear glasses or lenses. 4 (14.2\%) of the participants reported left handedness, while remaining reported right handedness. However, in the data collection, there were 16.5\% left handed pointing events, and 0.02\% pointing events that used both hands.

Due to a few administrative and technical reasons, the referencing events per user are not perfectly balanced. For the point based referencing (explained later in section \ref{sec:design}), only 18 of the 28 participants took part.  

\subsection{Experiment design}
\label{sec:design}

\begin{figure*}[t]
\centering
  \includegraphics[trim=0 0 0 0,clip, width=\textwidth]{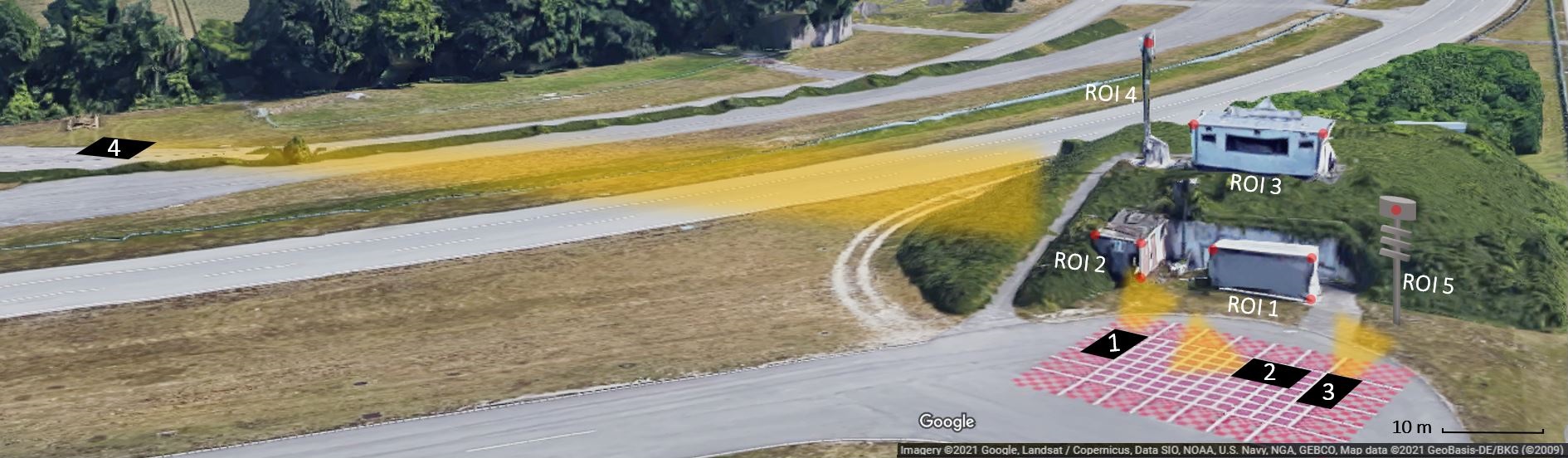}
  \caption{Map of the chosen 5 ROIs with black rectangles as the 4 car poses. The 10 red dots show the POIs.}
  \label{fig:map}
\end{figure*}

The environment\footnote{GPS coordinates 48.220418 N, 11.724965 E} was carefully chosen to have a large variety of pointing directions while having the landmarks relatively close to each in order to have a complex recognition scenario. The landmarks included small buildings and antennas, which are shown in Figure \ref{fig:map}. The participants were asked to point in a natural manner using the phrase, "what is \textit{that}?" with the pointing gesture. They were free to choose the hand for the pointing gesture so that the naturalness of the gesture is not restricted. 

The experiments consist of two types of referencing behavior : i) volume-based referencing and ii) point-based referencing. In volume-based referencing, the participants were told to point to a landmark, hereafter called Region of Interest (ROI), without any restrictions. This represents the natural and practical way of referencing as typically people reference objects that have a surface area. In contrast, in point-based referencing, the participants were told to point to a specific point on the landmark, which will be called Point of Interest (POI) hereafter. This referencing type allows precisely measuring the pointing. Both of the reference types were conducted from four different orientations or poses of the car (shown in Figure \ref{fig:map}) for a large variance. The first three poses consist of ROIs lying near the vehicle (i.e., 10m to 30m away), while the fourth pose consists of ROIs further away (more than 100m away). The first car pose had ROIs lying on both sides, the second pose was chosen such that all ROIs were situated on the right side, while the third pose was chosen to keep the majority of the ROIs on the left side. The positions of ROIs are categorized as being far left or far right if they lie more than 45° on the left or right side of the car, respectively, and near left or near right if they lie less than 45° to the left or right side, respectively. The details of each car pose for these categorizations are summed up in Table \ref{Table:poses}.

\begin{table}[h]
\centering
 \begin{tabular}{c | c c c c | c c }
 \hline
 \textbf{Car} & \textbf{Far} & \textbf{Near} & \textbf{Near} & \textbf{Far} & \textbf{Distance} & \textbf{Referencing}\\
 \textbf{pose} & \textbf{left} & \textbf{left} & \textbf{right} & \textbf{right} & &\textbf{type}\\
 \hline\hline 
 Pose 1 & & \checkmark & \checkmark & & near & point+volume\\
 \hline
 Pose 2 & \checkmark & \checkmark & & & near &  point+volume\\
 \hline
 Pose 3 & & & \checkmark & \checkmark & near &  point+volume\\
 \hline
 Pose 4 & & \checkmark &  \checkmark & & far & volume\\
 \hline
\end{tabular}
\caption{Categorization of ROIs based on their position.}
\label{Table:poses}
\end{table}

\subsection{Dataset statistics}
For the measured directions of each modality from the camera systems, the angular distance from the ground truth direction is calculated in the yaw (horizontal) and pitch (vertical) directions. The mean (M) and standard deviation (SD) of the angular distance for each modality are shown in Table \ref{Table:stats} w.r.t. the four car poses. These statistics include samples from all participants, where each sample consists of the entire 0.8 seconds. From amongst all car poses, pose 4 has the least deviation from the ground truth, as the driver mainly looks straight ahead. Eye direction has relatively small deviations from ground truth in pose 1. However, the large M and SD in pose 2 and 3 for both head and eye direction could be due to occlusion (in which case interpolation was used). Finger direction has large M and SD values in all poses. It was observed that the finger pointing lagged the gaze fixation by a small amount (details are omitted from this paper). Therefore, the time interval chosen involves a relatively large change in the finger direction, compared to head and eye, from the hands placement on the steering wheel to the respective pointing gesture. This change is largest in pose 2 as it has ROIs/POIs with the farthest angular distance (see Figure \ref{fig:map}). 

\begin{table}[t]
\centering
\resizebox{\columnwidth}{!}{
 \begin{tabular}{c|c c|c c|c c}
  \hline 
  & \multicolumn{2}{c|}{\textbf{Head Direction}} & \multicolumn{2}{c|}{\textbf{Eye Direction}} & \multicolumn{2}{c}{\textbf{Finger Direction}} \\
  
 \textbf{Car} & \textbf{Yaw} & \textbf{Pitch} & \textbf{Yaw} & \textbf{Pitch} & \textbf{Yaw} & \textbf{Pitch}  \\
 \textbf{pose}  & \textbf{M (SD)} & \textbf{M (SD)} & \textbf{M (SD)} & \textbf{M (SD)}& \textbf{M (SD)} & \textbf{M (SD)}   \\  
 \hline  \hline
 1 & 18° (13°) & 20° (15°) & \ 8° (10°) & 15° (11°) & 28° (24°) & 27° (17°) \\
 2 & 31° (21°) & 21° (18°) & 25° (25°) & 18° (11°)  & 49° (42°) & 19° (14°)\\
 3 & 25° (16°) & 16° (13°) & 14° (17°) & 16° (12°)  & 31° (32°) & 25° (16°)\\
 4 & \ 9° ( 9°) & \ 9° ( 7°) & \ 6° ( 7°) & 12° ( 9°)  & 24° (19°) & 25° (16°) \\
 All & 24° (18°) & 19° (15°) & 15° (19°) & 16° (12°) & 35° (34°) & 24° (16°) \\
  \hline
\end{tabular}}
\caption{Statistics of the angular distance of measured direction of modalities from the ground truth.}
\label{Table:stats}
\end{table}

\section{Multimodal Object Referencing Framework}

\subsection{Ground Truth Definition}
The GPS coordinates of all the vertices of the ROIs, and the ground contact point of all tyres for the car parked in the four different poses, were measured in  WGS84 (World Geodetic System) standard by a laser sensor, Leica Multistation\footnote{{https://leica-geosystems.com/en-us/products/total-stations/multistation}}.  The geodetic coordinates (latitude $\phi$, longitude $\lambda$ and altitude $h$) were converted to Earth-Centered, Earth-Fixed (ECEF) Cartesian coordinates ($x$ , $y$, $z$) using,

\begin{align}
    x &= (N(\phi) + h ) \ \text{cos} \phi \ \text{cos} \lambda \\
    y &= (N(\phi) + h ) \ \text{cos} \phi \ \text{sin} \lambda \\
    z &= (1 - e^2) (N(\phi) + h) \ \text{sin} \phi
\end{align}
where $N(\phi)$ is the prime vertical radius of curvature, and $e$ is the eccentricity of ellipsoid related to the semi-major axis, $a$, and the semi-minor axis, $b$, \footnote{ The semi-major and semi-minor axes are taken as, $a = 6378137.0 $ m and $ b = 6356752.3142 $ m, respectively.} as: 
\begin{align}
    N(\phi) & = \frac{a}{\sqrt{1 - e^2 \ \text{sin}^2\phi}} \ \ , \ \ \ \
    e^2 = 1- \frac{b^2}{a^2}
\end{align}

An affine transformation was applied to convert the ECEF cartesian coordinates to the car coordinate system, using the rotation and translation matrices calculated from the car pose.

Ground truth is then defined as a unit vector from origin towards the POI for point based referencing, and as a unit vector from origin towards the centre of the ROI (calculated as mean of all vertices) for volume based referencing. 
Notice that the ground truth is different for both referencing types.

\subsection{Fusion Model}
The overall architecture is shown in Figure \ref{fig:methodology}. The features from the camera systems were preprocessed, as mentioned in Section \ref{sec:feats}, and merged using a deep CNN in a model-level fusion approach. This supervised learning approach is used to cover a large number of different cases, which the CNN can learn on its own provided there is large variance in the dataset. As the features constitute a temporal sequence, one of the dimensions upon which the convolution block operates is the time dimension. This way, the model-level fusion also provides the CNN with the tendency to implicitly learn the temporal relations between modalities \cite{wu2014survey}.

A batch of samples of size $b$ was used as input, $x$, to the model, such that $x \in \mathbb {R}^{b \times t \times f \times d}$, where $t$ is the number of temporal (consecutive) frames, $f$ is the number of features and $d$ is the number of dimensions in each feature. Each sample consisted of $t=36$ consecutive frames ($\approx 0.8$ sec, at 45 frames per second) and $f=6$ features: the position and direction of the three modalities. Each of these were 3 dimensional, i.e., $d=3$.

The CNN model first applies two convolutional layer operations on each of the eye, head and finger poses separately totalling six layers, with each layer using kernel size of $(1 \times 1)$. Next, a concatenation layer is used to fuse all the feature maps and two convolutional layers are further applied with a kernel size of $(2 \times 2)$.  Rectified Linear Unit (ReLU) layers are used as activation after each of the convolutional layers. Each convolutional layer consists of 128 feature maps. A flatten layer is applied before the final dense layer, which uses linear activation to provide the linearly regressed fused 3D direction vector.
For the loss function, $\mathcal{L}$, the Mean Angular Distance (MAD) between the fused vector and the ground truth vector is used. In other words, the angle between the two vectors is minimized. Mathematically:

\begin{align}
    \mathcal{L} &= \text{MAD} = \frac{1}{N} \sum^N_{i=1}  \theta_i  \\
    &= \frac{1}{N} \sum^N_{i=1}  \text{arccos} \left( \frac{\textbf{\^y}_i \ . \ \textbf{y}_i}{\|\textbf{\^y}_i \| \  \| \textbf{y} _i\|} \right)  \ \ \ \in \ [-1, 1]
    \label{eqn}
\end{align}

where $ \textbf{ \^y}_i$ is the $i$-th predicted vector, $\textbf{y}_i$ is the $i$-th ground truth vector, $\theta_i$ is the angle between the two 3D vectors, and $N$ is the total number of samples.

\subsection{Matching Algorithm}
The referencing direction (as a normalized fused vector) predicted by the multimodal fusion model is matched with the ROI with which it has the minimum Euclidean distance, such that the distance is zero if the predicted vector lies within the boundaries of the ROI. If two or more ROIs have zero distance (for example when one ROI partially ocludes the other), then minimum cosine proximity of the fused vector to centre of the ROIs is used to predict the target ROI. The method is explained in Algorithm \ref{alg:1}.

\begin{algorithm}[t]
 \caption{Matching predicted vector to ROI}
\SetAlgoLined
    
    \SetKwInOut{Input}{input}
    \SetKwInOut{Output}{output}
    \SetKwProg{Init}{initialize :}{}{}

 \Input{Fused vector, $\textbf{v}_f = [v_x,v_y,v_z] \in \mathbb{R}^{1 \times 3}$ , Map of  ROIs $= [\textbf{p}_0, \textbf{p}_1, ..., \textbf{p}_{n-1}] \in \mathbb{R}^{n \times 8 \times 3}$, car\_pose $\in \mathbb{R}^{4 \times 3}$}
 \Init{}{
 $\hat{\textbf{v}}_f \gets \textbf{v}_f / \|\textbf{v}_f\|$\;
 $\textbf{d} = \left[ d_0, d_1, ... , d_{n-1} \right] = [\infty, \infty, ... , \infty ]$}  
 \For{i = 0 to (n-1)}
 {
  $\textbf{p}_{c} \gets$ WGS84\_to\_car\_coordinates ($\textbf{p}_i$, car\_pose)\;
  $\hat{\textbf{p}}_{c} \gets {\textbf{p}_{c}} / {\|\textbf{p}_{c}\|}$\; 
  
  $d_x \gets \text{max} \left(\text{min}_x(\hat{\textbf{p}}_{c}) - \hat{v}_x, \ 0, \ \hat{v}_x - \text{max}_x(\hat{\textbf{p}}_{c}) \right)$\;
  $d_y \gets \text{max} \left(\text{min}_y(\hat{\textbf{p}}_{c}) - \hat{v}_y, \ 0, \ \hat{v}_y - \text{max}_y(\hat{\textbf{p}}_{c}) \right)$\;
  $d_z \gets \text{max} \left(\text{min}_z(\hat{\textbf{p}}_{c}) - \hat{v}_z, \ 0, \ \hat{v}_z - \text{max}_z(\hat{\textbf{p}}_{c}) \right)$\;
  $d_i \gets \sqrt{(d_x)^2 + (d_y)^2 + (d_z)^2}$ \;
 }
\uIf{$\text{count}(d_i == 0) > 1 \ \ \forall \ i \in [0, 1, ... , n-1]$}{
    indices $\gets$ where($d == 0$) \;
    roi $\gets$ argmin(cosine\_proximity (mean($\textbf{p}_{j}), \textbf{v}_f$)) \ $\forall \ j \in [\text{indices}]$
  }
  \Else{
    roi $\gets$ argmin($d_0, d_1, ..., d_{n-1}$)\;
  }  

  \Output{roi}
\label{alg:1}
\end{algorithm}

\section{Experiments and Results} 
We used 5-fold cross validation to evaluate our approach on the collected dataset.
The entire data is split into training, validation and test sets. Since the data is unbalanced among participants, we chose the splits to be user-based to avoid user specific bias, i.e., no reference sample from participants in the training set appears in either the validation or the test set. Each test set contains five or six users and approximately the same number of samples. For each experiment, the CNN model was trained for 50 epochs using a batch size of 32 and the Adam optimizer with variable learning rate starting from 0.001. The model which minimizes the validation loss is used to evaluate the test set. The metrics are averaged over the 5 folds in the test sets to incorporate the entire data.

\subsection{Metrics}
\subsubsection{Accuracy metrics}
Our main goal is to be able to correctly identify (i.e. classify) the referenced ROIs. Therefore, we use a 5-class accuracy metric for this:
\begin{align}
    \text{Accuracy} = \frac{TP}{{N}} \times 100 \%
\end{align}
where $N$ is the total number of predictions and $TP$ is the total number of true predictions (or correct identification) by the model. 
We also use the top-2 accuracy, to see the percentage of the true ROI being among the top two predicted candidates, that may be used when a list of candidate ROIs is presented to the driver to choose the intended one.  

\subsubsection{Mean Angular Distance (MAD)} MAD is the mean of the angles between the predicted vectors and the corresponding ground truth vectors. It used to evaluate the precision of the regression output. Consequently, the lower the MAD, the higher the precision. The standard deviation of the angular distance (Std.AD) is also used to measure the variation of the predictions. It is important to note that ground truth is different for the two types of referencing.

\begin{table}[t]
\centering
\resizebox{\columnwidth}{!}{
 \begin{tabular}{c c c|c c c}
 \hline
 \textbf{Car} & \textbf{\# of} & \multirow{2}{*}{\textbf{Modality}}  & \multirow{2}{*}{\textbf{Acc.} $\uparrow$} & \textbf{Top$-2$ }  & \multirow{2}{*}{\textbf{MAD (Std.AD)} $\downarrow$ }\\  
 \textbf{Pose} & \textbf{events} & & & \textbf{Acc.} $\uparrow$ & \\
 \hline\hline
 \multirow{4}{*}{Pose 1} & \multirow{4}{*}{2,958} & Head & 51.5\% & 76.2\% & 13.0° \ (11.7°)  \\
  & & Gaze & 86.7\% & 96.3\% & \ 4.0° \ ( 6.7°)  \\
  & & Finger  & 67.3\% & 83.0\% & \ 9.0° \ (11.3°)  \\
  & & \textbf{Fusion} & \textbf{90.0\%} & \textbf{98.0\%} & \textbf{ 3.4° \ ( 4.8°)}  \\ 
 \hline
 \multirow{4}{*}{Pose 2} & \multirow{4}{*}{2,871} & Head  & 49.1\% & 70.5\% & 12.5° \ (11.8°)  \\
  & & Gaze    & 48.8\% & 65.6\% & 14.6° \ (14.2°)  \\
  & & Finger  & 61.6\% & 78.6\% & \ 9.8° \ (10.7°)  \\
  & & \textbf{Fusion} & \textbf{75.6\%} & \textbf{87.7\%} & \textbf{ 7.3° \  ( 8.9°)}  \\
 \hline
 \multirow{4}{*}{Pose 3} & \multirow{4}{*}{2,830} & Head  & 46.6\% & 67.4\% & 10.7° \ ( 9.3°)  \\
  & & Gaze    & 66.4\% & 83.7\% & \ 6.1° \ ( 6.7°)  \\
  & & Finger  & 27.3\% & 49.1\% & 14.4° \ ( 9.8°)  \\
  & & \textbf{Fusion} & \textbf{68.4\%} & \textbf{84.2\%} & \textbf{ 5.8° \ ( 5.7°)}  \\
 \hline
 \multirow{4}{*}{Pose 4} & \multirow{4}{*}{1,502} & Head  & 31.3\% & 61.2\% & \ 3.9° \ ( 3.1°)  \\
  & & \textbf{Gaze}    & \textbf{65.2\%} & \textbf{89.9\%} & \textbf{ 1.4° \ ( 1.9°)}  \\
  & & Finger  & 46.2\% & 74.3\% & \ 2.6° \ ( 2.2°)  \\
  & & Fusion & {57.9\%} & {89.2\%} & \ 1.5° \ (\ 1.7°)  \\
 \hline
  & \multirow{4}{*}{10,161} & Head  & 44.3\% & 69.4\% & 15.1° \ (14.0°)  \\
  All & & Gaze    & 65.4\% & 82.1\% & \ 9.5° \ (13.7°)  \\
  poses & & Finger  & 45.5\% & 69.1\% & 17.2° \ (22.4°)  \\
  & & \textbf{Fusion} & \textbf{72.2\%} & \textbf{88.5\%} & \textbf{ 6.8° \ ( 8.5°)}  \\
 \hline
\end{tabular}}
\caption{Volume based referencing.}
\label{Table:volume_results}
\end{table}

\subsection{Ablation study}
\label{sec:ablation}
We started with an ablation study to see the effects of each individual modality for every car pose. Each car pose contains the ROIs in a different setting (see Table \ref{Table:poses}). Tables \ref{Table:volume_results} and \ref{Table:point_results} show the detailed results for the volume-based referencing and point-based referencing, respectively. It is worth noting that for the distant objects, only three of the ROIs were visible but the POIs were not visibly distinct to the drivers. Therefore, pose 4 (with distant objects) is not included in the point-based referencing but is still included in volume-based referencing. However, it contains fewer number of reference events as compared to the rest due to less number of ROIs. Nevertheless, matching of the predicted vector to a ROI is done while considering all five ROIs.

For each experiment setting (i.e. each row) in the tables, a new trained model is used, e.g. for head modality and pose 1, model is trained only using head data from the car situated in pose 1. The best values of each metric in each pose are shown in bold in the tables. It can be seen from both tables \ref{Table:volume_results} and \ref{Table:point_results} that, in all cases, the fusion of the three modalities (head, gaze and finger) results in a higher accuracy.

\begin{table}[t]
\centering
 \resizebox{\columnwidth}{!}{
 \begin{tabular}{c c c | c c c}
 \hline
 \textbf{Car} & \textbf{\# of} & \multirow{2}{*}{\textbf{Modality}}  & \multirow{2}{*}{\textbf{Acc.} $\uparrow$} & \textbf{Top$-2$ }  & \multirow{2}{*}{\textbf{MAD (Std.AD)} $\downarrow$ } \\  
 \textbf{Pose} & \textbf{events} & & & \textbf{Acc.} $\uparrow$ & \\
 \hline\hline
 \multirow{4}{*}{Pose 1} & \multirow{4}{*}{4,776} & Head  & 49.9\% & 78.5\% & 13.9° \ (10.9°)  \\
  & & Gaze    & 78.3\% & 96.3\% &  4.4° \ ( 6.3°)  \\
  & & Finger  & 57.5\% & 82.5\% & 11.4° \ (10.5°)  \\
  & & \textbf{Fusion} & \textbf{79.1\%} & \textbf{96.5\%} &  \ \textbf{4.4° \ ( 5.2°)}  \\
 \hline
 \multirow{4}{*}{Pose 2} & \multirow{4}{*}{4,206} & Head  & 55.6\% & 77.3\% & 12.3° \ ( 9.0°)  \\
  & & Gaze    & 64.5\% & 82.0\% & 10.4° \ ( 9.2°)  \\
  & & Finger  & 68.4\% & 88.9\% & \ 8.6° \ ( 7.4°)  \\
  & & \textbf{Fusion} & \textbf{74.7\%} & \textbf{92.7\%} & \textbf{ 6.4° \ ( 5.5°)}  \\ 
 \hline
 \multirow{4}{*}{Pose 3} & \multirow{4}{*}{4,648} & Head  & 50.9\% & 76.2\% & 12.7° \ (10.2°)  \\
  & & Gaze    & 68.8\% &  91.1\% & \textbf{ 6.5° }\ ( 6.4°) \\
  & & Finger  & 31.0\% & 58.3\% & 16.9° \ (10.2°)  \\
  & & \textbf{Fusion} & \textbf{70.0\%} & \textbf{91.1\%} & \ 6.6° \ \textbf{( 5.9°)}  \\
 \hline
  & \multirow{4}{*}{13,630} & Head  & 50.9\% & 76.4\% & 16.4° \ (13.4°)  \\
  All & & Gaze    & 68.5\% & 89.2\% & \ 8.5° \ (10.2°)  \\
  poses & & Finger  & 49.6\% & 72.3\% & 18.3° \ (20.4°)  \\
  & & \textbf{Fusion} & \textbf{73.6\%} & \textbf{92.5\%} & \textbf{ 6.7° \ ( 7.5°)}  \\
 \hline
\end{tabular}}
\caption{Point based referencing.}
\label{Table:point_results}
\end{table}

\subsubsection{Effect of modalities}
Amongst the three modalities, we observe gaze to be most accurate for both volume-based and point-based referencing. For vehicle pose 1 and 4 (which are most representative of real everyday driving on the road), gaze achieves a MAD of about 4° and 1.4°, respectively,  which are significantly better than the other two modalities. Similarly, the classification accuracy of gaze in all poses except pose 2 is notably higher than head and finger modalities. The reason why performance of gaze is weaker in pose 2 is explained later in section \ref{sec_pose2}

Head pose as a single modality does not provide much information to distinguish closely situated objects. Finger direction appears to have relatively large deviations compared to gaze, but assists gaze slightly when available. The results in Tables \ref{Table:volume_results} and \ref{Table:point_results} reveal that in almost all cases multimodal fusion is more precise (i.e. lower MAD), and it offers higher recognition accuracy than monomodal approaches such as those presented in \cite{rumelin2013free, kang2015you}. 

\subsection{Limitations imposed by vehicle pose and reference type}
Volume-based referencing represents real world pointing behavior. However, we do not know the exact point of interest that was actually referenced. On the contrary, in point-based referencing, there is tendency for users to be relatively more precise but it only represents a small fraction of the real world behavior, since most landmarks that the driver could refer to would be large in size. The key points are summarized below:
\subsubsection{Pose 1} 
 For this case, a ROI recognition accuracy of 90\% is achieved with the true ROI lying among top-2 candidates 98\% of the time when users pointed freely to the volume. It is intuitive that accuracy for the point based referencing (79\%) comes out to be less, compared to volume based referencing, as users attempt to point to vertices of the ROIs which could be far from the centre of the ROIs. However, the higher MAD of 4.4° in point-based case compared to 3.4° in the volume-based case is counter intuitive, as one may hypothesize that users may attempt to be more precise in the former case. We observe that users tend to behave very differently when asked to point to precise points, evident from the very low R-squared value of 4.7\% of the linear regression of pose 1 in Figure \ref{fig:acc_v_mad_point}.

 While pose 1 is a typical referencing case, as the ROIs are situated in front and on the near sides only, pose 2 and 3 provide added limitations.

\subsubsection{Pose 2} \label{sec_pose2}
This is a relatively challenging referencing case as objects are situated on the far right side. For both point and volume based referencing types, finger pointing appears to be the most accurate modality as well as the most precise (i.e., lower MAD) (see Tables \ref{Table:volume_results} and \ref{Table:point_results}). 

In pose 2, when objects lie on the far right, drivers need to turn their head resulting in their eyes being out of the FoV of the VCS. The missing data is interpolated linearly but it is not enough to accurately predict the target ROIs. Head pose is still tracked by the VCS but as a single modality, is not of much use. However, fusion of all modalities, significantly improves both the accuracy as well as the precision.

\subsubsection{Pose 3}
In vehicle pose 3, the challenge is to recognize referencing to the far left side. Because of the placement of the GCS in the centre of the roof top, estimating gestures to left side is considerably difficult. As a result, finger pointing becomes the worst modality for object referencing in all metrics for pose 3 in both referencing types. 

The reasons behind these effects became clear when we looked deep into the sensory data and the recorded videos. In pose 3, with all ROIs situated on the left side of the driver, we observed 30\% of pointing events with the left hand. In all other poses, the left hand use is less than 10\%. Two phenomena happen in this pose. Firstly, the left hand was difficult to track by the GCS as left hand would lie near the edge of the FOV fish-eye, and often go beyond it. Occasionally, the right hand would also go beyond the FoV when users fully extended their arms to the left side. Secondly, with the right hand pointing to the left side, the eyes and face would become occluded by the arm. This resulted in missing gaze and head pose sensor data in some cases. Because of the incorrect tracking of the pointing gesture, fusion provides only has an insignificant enhancement compared to using gaze only.

\subsubsection{Pose 4} 
In pose 4, the objects were far in distance, and had small variations in the angular distances. This results in a very high precision. In this case, we see gaze to be the most precise with a MAD of 1.4°, while the fusion did not enhance it. The highest accuracy of 65\% is achieved using gaze only because slight movements of finger result in large angular deviations which deteriorate the fusion results.
\begin{figure}[t]
    \centering
    \includegraphics[width=\linewidth]{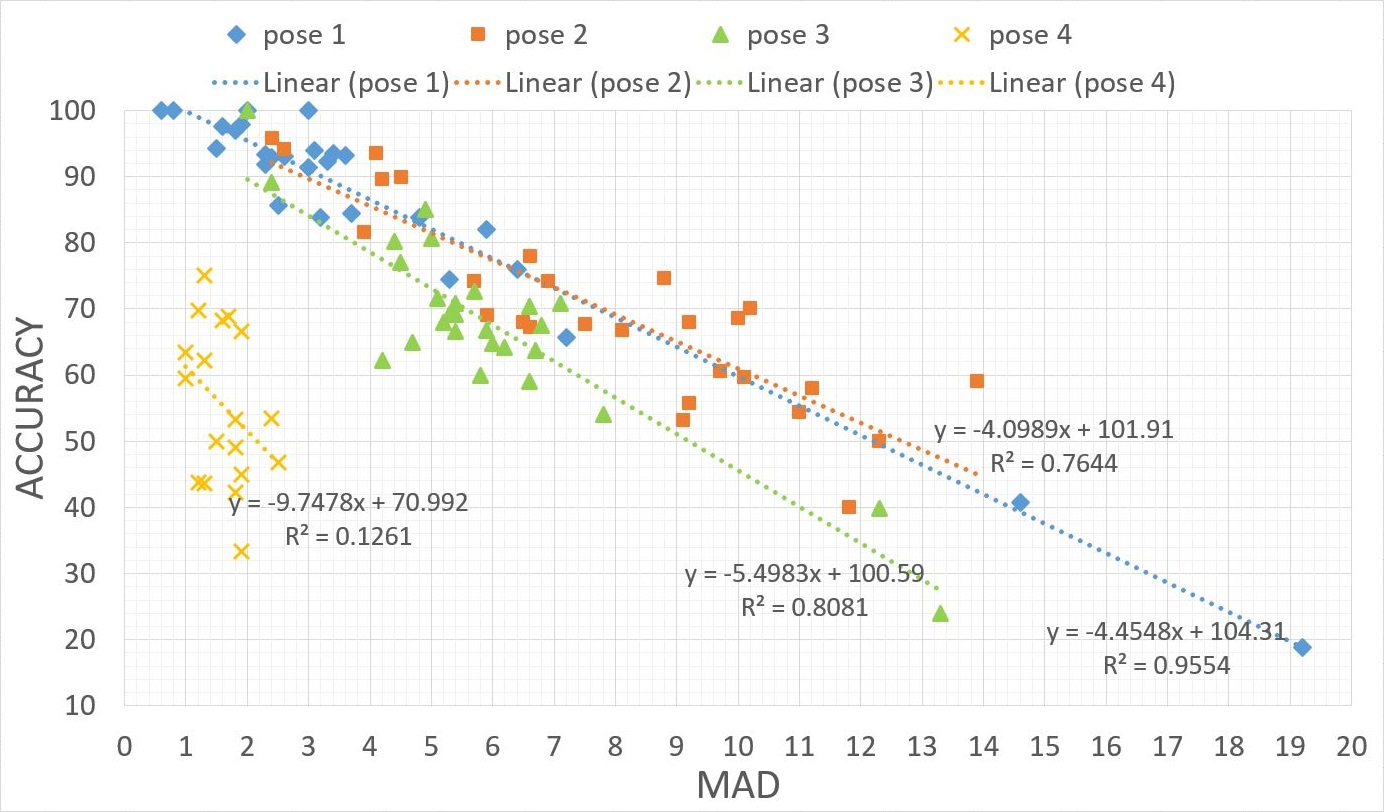}
    \caption{Volume-based user specific referencing performance.}
    \label{fig:acc_v_mad}
\end{figure}

\subsection{Combined training on all poses} 
The motivation for a combined training is to have one common model to cover all the objects that could be in the driver's view. Such a generalized model can be used for all cases (typical referencing as well as challenging ones). Intuitively, in both referencing types, gaze is most precise for predicting ROIs with an accuracy of 65\% and 69\% in the volume and point based referencing, respectively, while the fusion of all modalities further enhances this to 72\% and 74\%, respectively. The MAD in both types is approximately $7$° when all modalities are fused, which is an increase of about $2$° compared to gaze modality alone.


\begin{figure}[t]
    \centering
    \includegraphics[width=\linewidth]{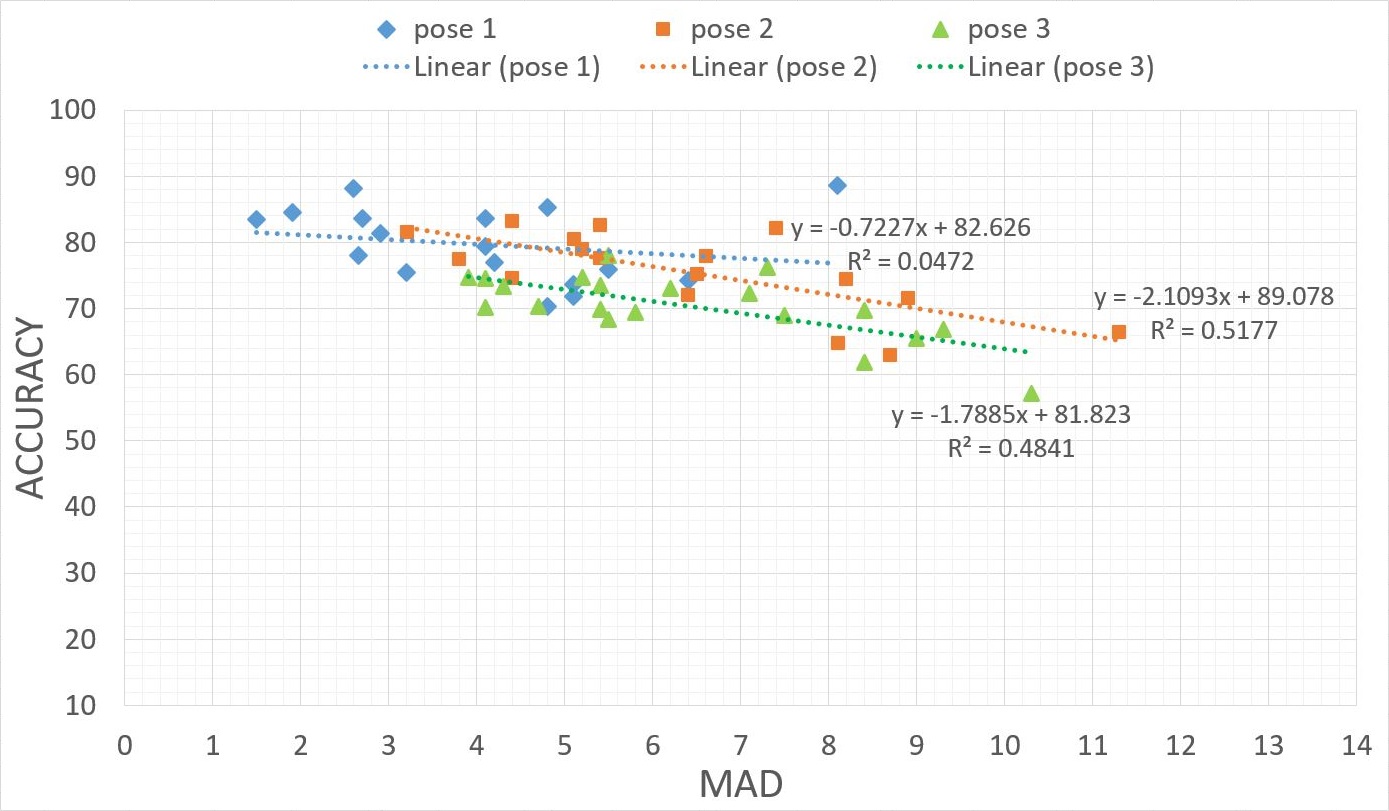}
    \caption{Point-based user specific referencing performance.}
    \label{fig:acc_v_mad_point}
\end{figure}

\subsection{Driver specific behavior}
A leave-one-out (28-fold) cross validation is used to obtain the driver specific results. The behavior of the drivers has a lot of variation between drivers as well as between poses. Scatter plots are shown in Figures \ref{fig:acc_v_mad} and \ref{fig:acc_v_mad_point} for volume and point based referencing, respectively. Each point represents the performance of the model (accuracy and MAD) for each driver, shown in different colors for the different poses. The linear regression fit for each pose is also plotted.

For volume based referencing, Figure \ref{fig:acc_v_mad}, we see a clear inverse relation between accuracy and MAD. Pose 1 (in blue) has a few outliers. The reason for the user with 19\% accuracy and 19° MAD is not entirely clear but is hypothesized (from examining the recorded videos) to be erroneous finger tracking as well as occlusion of face. The same user had better accuracy ($> 60\%$) and MAD ($< 7$°) in other poses.

It can be seen from Figure \ref{fig:acc_v_mad_point} that in point based referencing, the user behavior has a large variance (in MAD) for all poses but not in accuracy. All 3 poses have significantly smaller R-squared linear regression values for point based referencing compared to volume based referencing. It is likely that imposition of the restriction to precisely point, led to users overdoing the pointing gesture, resulting in different user behavior.


\section{Conclusion}
We extend our previous work \cite{aftab2020you} from the inside car use case to referencing objects outside the car, without the need of any touch based input. As compared to the previous work, this work contains a much larger dataset from 28 participants to have substantial variance for better learning of the deep CNN model. We conducted experiments in four different stationary car orientations for two different types of referencing to objects situated in the field-of-view of the driver. We conclude that the placement of sensors plays an important role when dealing with individual modalities. When using natural free hand pointing for referencing, occlusion of the face is very likely, which hinders the gaze and head pose tracking. Similarly, finger could go out of the FoV of the gesture tracker, which significantly effects it's performance. However, we show that using a deep CNN multimodal fusion architecture, the sensor based shortcomings can be overcome to predict the direction in which the driver referenced. We achieve a recognition accuracy of 90\% in the typical cases (when sensor data contains no or minimal occlusion), and an accuracy of more than 65\% 
in the most difficult cases that have missing sensor data (mainly due to occlusion while tracking the individual modalities). 

Furthermore, we observed the user behavior in two referencing types as well as in different environments, and discuss the differences in user referencing behavior. These findings motivate us to look into personalized multimodal models in the future for recognizing user specific behavior and improving predictions of the target object. The unique approach presented in this paper is also intended to provide considerable use of multimodal fusion for practical applications in the driving scenario for intelligent driver vehicle interaction, which we plan to investigate as well. 


\section*{Acknowledgements}
We are very grateful to Ovidiu Bulzan and Stefan Schubert (BMW Group, Munich) for their contributions to the experiment design and apparatus setup, and to Simon Stoeferle, Tobias Brosch and Wolfgang Mader (BMW Car IT GmbH, Ulm) for their support in data extraction and synchronization.



\bibliographystyle{ieeetr}
\bibliography{root}

\end{document}